\documentstyle{proceeding}
\input epsf.tex
\def\ginga{{\it Ginga}}
\def\g{$\gamma$}
\begin{document}
\title{The first simultaneous X-ray/$\gamma$-ray observations of Cyg X-1 
by {\it Ginga\/} and OSSE}
\author{Marek Gierli\' nski\inst{1}, Andrzej A. Zdziarski\inst{2}, W. Neil Johnson\inst{3},
Bernard F. Phlips\inst{3}, Ken Ebisawa\inst{4}, and Chris Done\inst{5}}  
\institute{Astronomical Observatory, Jagiellonian University, Cracow, Poland
\and Copernicus Astronomical Center, Warsaw, Poland
\and E. O. Hulburt Center for Space Research, Naval Research Lab, 
Washington DC, U.S.A.
\and NASA/GSFC, Greenbelt, U.S.A.
\and Department of Physics, University of Durham, Durham, U.K.}
\maketitle
\begin{abstract}

We present the results of 4 simultaneous observations of Cygnus X-1
by {\it Ginga} and OSSE. The X-ray/$\gamma$-ray spectra can
be described by an intrinsic continuum and a 
component due to Compton reflection including an iron K$\alpha$ line. 
The intrinsic spectrum at X-ray energies is a power-law with a photon
spectral index of $\Gamma=1.6$. The intrinsic $\gamma$-ray spectrum
can be phenomenologically described by either a power-law without 
cutoff up to 150 keV, and 
an exponential cutoff above that energy, or by an exponentially cutoff power
law {\it and\/} a second hard component.

\end{abstract}

\section{Introduction}

Cygnus X-1 is a binary system that consists of an O9.7 supergiant and 
a compact object which is believed to be a black hole. Its X-ray and
\g-ray spectrum shows enormous variability: the total flux can change
by a factor three within a few hours. On June 6, 1991 Cyg X-1 was observed
by \ginga\/ and {\it GRO}/OSSE simultaneously. This gives us a unique
opportunity to see instantaneous, wide-band spectra of Cyg X-1, 
covering the energy range from 2 to 1000 keV (Gierli\'nski et al. 1996).

\section{The spectra}

Fig. 1 shows four spectra of Cyg X-1, taken at different hours of June 6,
1991. The solid curves represent the two-component model described below.
The strongest and the weakest spectra differ in time by $\sim 15$ hours; 
the flux in the range 2--1000 keV had changed during this time by a factor 
$\sim 3$. The observed spectra fall into the high-state of Cyg X-1 as observed
in the OSSE energy range (Phlips et al.\ 1995).

\begin{figure}
\begin{center}
\leavevmode
\epsfxsize=8cm \epsfbox{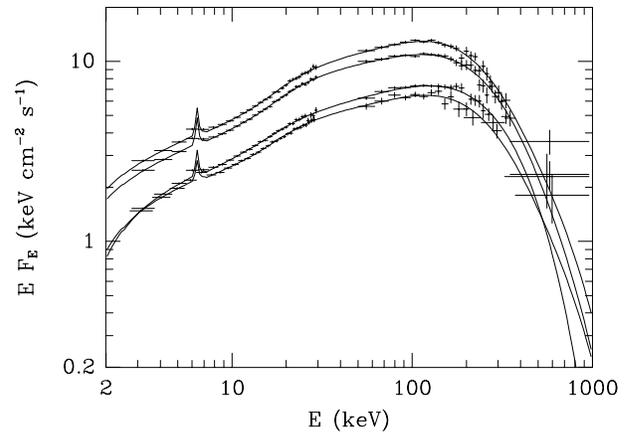}
\end{center}
\caption{
The four spectra of Cyg X-1 observed simultaneously by \ginga\ and OSSE
on June 6, 1991 between $0^{h}02^{m}$ and $20^{h}44^{m}$ UT. The spectra
correspond to data sets 2, 1, 3, 4, from top to bottom, respectively.
Solid curves 
represent the two-component model described in the text.}
\end{figure}

\section{\ginga\ and OSSE separately}

The \ginga\ data can be described well by a power-law intrinsic spectrum
accompanied by Compton reflection, which component includes a fluorescence
Fe K$\alpha$ at 6.4 keV.
The photon spectral index is $\sim 1.6$ and the covering factor of the 
reflecting medium (with an Fe overabundance of $\sim 1.5$) is 
$\sim 0.4$. We used angle-dependent reflection Green's functions 
(Magdziarz and Zdziarski 1995) and assumed the disk inclination angle of $30^o$.

The OSSE data are well described by either an exponentially cut-off 
power-law or a thermal Comptonization model. The photon spectral 
index is $\sim 1.0$, the $e$-folding energy, $E_f \approx 140$ keV, 
and the plasma temperature, $kT \approx 90$ keV.

\section{How to fit the joint \ginga\ and OSSE data?}

Although we find excellent fits for the \ginga\ and the OSSE data separately,
those fits do {\it not\/} match. We have found
that good fits to the joint 
\ginga\ and OSSE data require a modification of the form of the \g-ray cutoff
as well as an increase of the relative normalization of the \ginga\
data by $\sim 15$\%.
Two models providing good fits to the joint data are described below.

\subsection{Model one: exponential cutoff above 150 keV}

This model consists of a power-law with an exponential cutoff that
acts only above some cut-off energy $E_c$:

$$F_E = A E^{-\Gamma} \cases{ \exp(- (E - E_c) / E_f),&for $E > E_c$,\cr
                              1,&for $E \le E_c$.\cr}$$

The model includes Compton reflection continuum and an Fe line at 6.4 keV.
Fig.\ 2 shows how this model fits the data.  
We have found the photon index of $\Gamma \simeq 1.6$, the cut-off 
energy, $E_c \approx 150$ keV, and the $e$-folding energy, $E_f$, between 
200 and 240 keV for the four data sets.

\begin{figure}
\begin{center}
\leavevmode
\epsfxsize=8cm \epsfbox{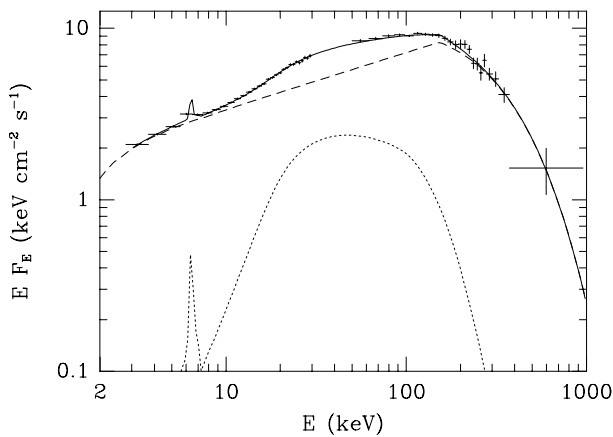}
\end{center}
\caption{
The power-law model exponentially cut-off above $\sim 150$ keV
for the \ginga/OSSE data set number 1. Angle-dependent disk reflection 
is shown with the dotted curve. The dashed curve represents the model 
without disk reflection. Best fit parameters for this model are: the photon spectral index $\Gamma = 1.67$, the cut-off energy
$E_c = 152$ keV, the $e$-folding energy $E_f = 205$ keV, the covering factor of the 
reflector, $\Omega/2 \pi = 0.40$, and $\chi^2 = 48$ (74 d.o.f.)}
\end{figure}

\subsection{Model two: two power-law components}

The base for this model is a simple, exponentially cut-off power-law
with Compton reflection and a Gaussian line at 6.4 keV. However, this model
requires an addition of another component peaking around 200 keV, 
represented by an exponentially cut-off ($E_f \sim 70$ keV), 
hard power-law ($\Gamma < 0$). Fig. 3 shows the resulting fit.

\begin{figure}
\begin{center}
\leavevmode
\epsfxsize=8cm \epsfbox{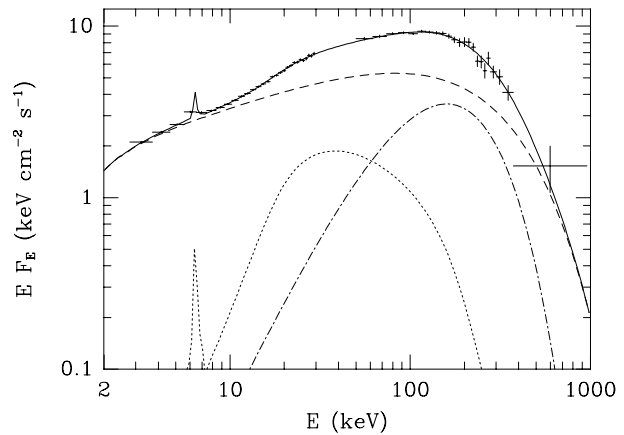}
\end{center}
\caption{
The two-component model for the \ginga/OSSE data set number 1.
The dashed curve represents the main, exponentially cut-off power-law 
component, the dotted curve---a disk reflection continuum, and the dash-dotted 
curve---the additional 
cut-off power-law component. Solid curve shows the sum.
Best fit parameters are: the photon spectral
index $\Gamma = 1.62$, the $e$-folding energy, $E_f = 217$ keV,
the covering factor of the reflector, $\Omega/2 \pi = 0.38$. The second component 
parameters are: the photon spectral index $\Gamma' = -0.23$ and the $e$-folding
energy, $E_{c}' = 72$ keV. $\chi^2 = 45$ (74 d.o.f.)}
\end{figure}

\section{Discussion}

Our results demonstrate unambiguously the presence of Compton reflection in 
Cyg X-1. The reflection continuum is accompanied by a Fe K$\alpha$ fluorescent
line with an equivalent width of about 100 eV, as expected theoretically
(George \& Fabian 1991). The observed spectra are cut off above $\sim 150$ keV;
however the form of the cutoffs is described neither by an exponentially cut
off power law nor by thermal Comptonization (treated relativistically). This 
effect can be due to a distribution of the plasma temperature and optical depth.
Alternatively, the cutoff can be reproduced by adding a hard component
peaking at 200 keV.
Note that the spectral index of the X-ray power
law, $\Gamma\sim 1.6$, is significantly harder than the average spectral 
index in Seyfert 1's ($\Gamma\sim 1.9$, Nandra \& Pounds 1994).

\end{document}